# Design of silicon waveguides for Kerr nonlinear optical performance with graphene oxide films

Yuning Zhang, Jiayang Wu, *Member, IEEE*, Yang Qu, Linnan Jia, Baohua Jia, *Fellow, OSA*
and David J. Moss, *Fellow, IEEE*, *Fellow, OSA*

*Abstract*—The Kerr nonlinear optical performance of silicon nanowire waveguides integrated with 2D layered graphene oxide (GO) films is theoretically studied and optimized based on experimentally measured linear and nonlinear optical parameters of the GO films. The strong mode overlap between the silicon nanowires and highly nonlinear GO films yields a significantly enhanced Kerr nonlinearity for the hybrid waveguides. A detailed analysis for the influence of waveguide geometry and GO film thickness on the propagation loss, nonlinear parameter, and nonlinear figure of merit (FOM) is performed. The results show that the effective nonlinear parameter and nonlinear FOM can be increased by up to ~52 and ~79 times relative to bare silicon nanowires, respectively. Self-phase modulation (SPM)-induced spectral broadening of optical pulses is used as a benchmark to evaluate the nonlinear performance, examining the trade-off between enhancing Kerr nonlinearity and minimizing loss. By optimizing the device parameters to balance this, a high spectral broadening factor of 27.8 can be achieved – more than 6 times that achieved in previous experiments. Finally, the influence of pulse chirp, material anisotropy, and the interplay between saturable absorption and SPM is also discussed, together with the comparison between the spectral broadening after going through GO-coated and graphene-coated silicon waveguides. These results provide useful guidance for optimizing the Kerr nonlinear optical performance of silicon waveguides integrated with 2D layered GO films.

## I. INTRODUCTION

Third-order nonlinear optical processes, including self-phase modulation (SPM), cross phase modulation (XPM), four-wave mixing (FWM), third harmonic generation (THG), and others [1, 2], have formed the basis for all-optical signal generation and processing, which can achieve ultrahigh processing speed without the need to convert the optical signal to the electrical domain (or vice versa) [3-5]. This has underpinned many applications in telecommunications [6], metrology [7], astronomy [8], ultrafast optics [9], quantum photonics [10], and others [11-13]. As an important third-order nonlinear optical process arising from Kerr effect, SPM that occurs when an optical pulse with a high peak power propagates through a nonlinear medium has been widely utilized for broadband optical sources, pulse compression, optical spectroscopy, and optical coherence tomography [12, 14].

Realizing Kerr nonlinear optical devices in integrated photonic chips would reap the greatest benefits in terms of device footprint, scalability, stability, and mass production. Although silicon has been a dominant platform for integrated photonic chips [15-20], its strong two-photon absorption (TPA) in the near-infrared telecom wavelength band significantly limits its nonlinear performance [1, 2]. To mitigate this challenge, p-i-n junctions across the silicon waveguides have been introduced to sweep the free carriers generated by TPA, which have greatly improved the response speed and have enabled many silicon electro-optic modulators with notable modulation speed [21-24]. However, while p-i-n junctions mitigate the free carrier effects very effectively, they do not increase the intrinsic nonlinear figure of merit (FOM) of silicon, which is fundamental and determined solely by the bandgap. The relatively poor nonlinear FOM of silicon in the telecom band is still much less than what is needed to achieve superior nonlinear performance. In response to this, other complementary metal-oxide-semiconductor (CMOS) compatible platforms have been explored for nonlinear optics, such as silicon nitride (SiN) [25, 26] and Hydex [27-79]. However, while these platforms have negligible TPA, they also have a comparatively low Kerr nonlinearity [2, 29].

To overcome these limitations, two-dimensional (2D) materials that exhibit an ultrahigh optical nonlinearity, such as graphene [80, 81], graphene oxide (GO) [82, 83], black phosphorus [84, 85], and transition metal dichalcogenides (TMDCs) [86, 87], have been integrated onto chips to enhance the nonlinear optical performance. Amongst the different 2D materials, GO has become highly promising due to its ease of preparation as well as the flexibility in tuning its material properties [88-94]. Previously, GO has been shown to have a giant Kerr nonlinearity – about 4 orders of magnitude higher than silicon [92, 95]. Moreover, GO has a large bandgap (> 2 eV [88, 96]) that yields a linear absorption that is over 2 orders of magnitude lower than undoped graphene at infrared wavelengths [97] as well as low TPA in the telecom band [96, 98]. Based on this, enhanced SPM in GO-coated silicon nanowires [82] and FWM in GO-coated SiN and Hydex devices have all been demonstrated [83, 97, 99]. An even more appealing advantage of GO is the ability to precisely control the film thickness, size, and position on integrated chips via large-area, transfer-free, layer-by-layer coating methods together with standard lithography and lift-off processes [96, 99, 100]. In contrast to the imprecise, largely unrepeatable, and unstable approach of mechanical layer transfer processes that have been widely used for other 2D materials such as graphene and TMDCs [101, 102], this method enables cost-effective, large-scale, and highly precise integration of 2D layered GO films on a chip, representing a significant advance towards manufacturable integrated photonic devices incorporating 2D materials [94].

Recently [82], we demonstrated an enhanced Kerr nonlinearity in silicon-on-insulator (SOI) nanowires integrated with 2D layered GO films, verified through SPM measurements with picosecond optical pulses. We achieved a maximum spectral broadening factor (BF) of 4.34 for an SOI nanowire with a patterned GO film. In this paper, we fully analyze and optimize the Kerr nonlinear optical performance of GO-coated SOI nanowires based on experimentally measured linear and nonlinear optical parameters of the GO films. We investigate the influence of waveguide geometry and GO film thickness on the propagation loss, nonlinear parameter, and nonlinear FOM. By increasing the mode overlap with GO films, we show that the effective nonlinear parameter and nonlinear FOM of the hybrid waveguides can be increased by up to ~52 and ~79 times with respect to bare SOI nanowires, respectively. We find that this needs to be balanced with an accompanying increase in linear loss, and use SPM-induced spectral broadening of the optical pulses to examine the trade-off between enhancing the Kerr nonlinearity and minimizing loss. By changing the device parameters to balance this trade-off, we achieve a high spectral BF of 27.8, more than 6 times higher than what has been achieved experimentally. Finally, we discuss the influence of pulse chirp, material anisotropy, and the interplay between saturable absorption (SA) and SPM on the Kerr nonlinear optical performance and compare the spectral broadening after passing GO-coated and graphene-coated SOI nanowires. These results highlight the significant potential to improve on experimental results [82] and provide detailed solutions for optimizing the Kerr nonlinear performance of SOI nanowires integrated with 2D layered GO films.

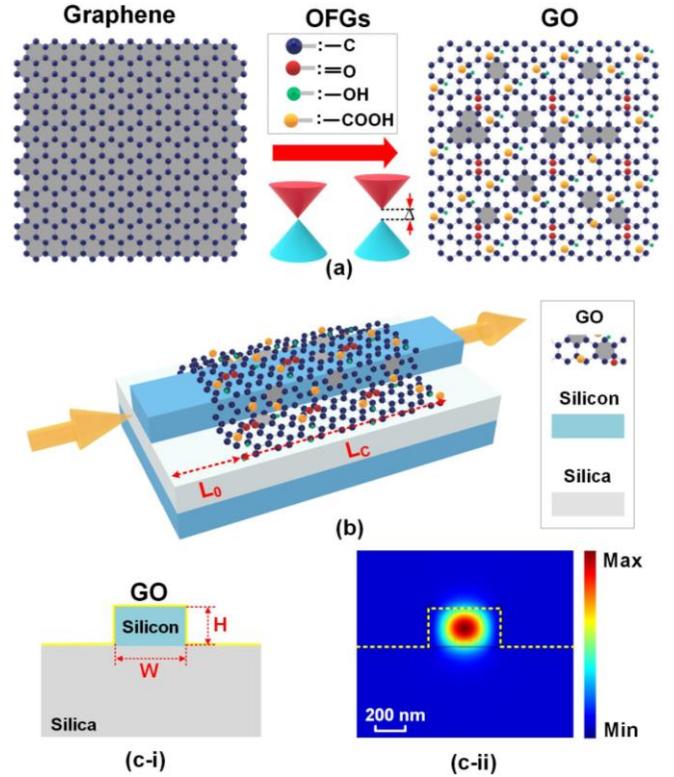

Fig. 1. (a) Schematics of atomic structures and bandgaps of graphene and GO. (b) Schematic illustration of a GO-coated SOI nanowire. (c-i) Schematic illustration of cross section of a SOI nanowire conformally coated with 1 layer of GO. (c-ii) TE mode profile corresponding to (c-i). The definitions of $L_0$, $L_c$, $W$, and $H$ are given in Table I.  OFG: oxygen-containing functional groups

## II. 2D GO FILMS AND DEVICE STRUCTURE

Fig. 1(a) shows schematics of the atomic structures and bandgaps of graphene and GO. As compared with graphene, GO provides more flexibility to tailor its material properties by manipulation of the oxygen-containing functional groups (OFGs) in the basal plane and sheet edges, including epoxy, hydroxyl, carbonyl and carboxyl groups [88, 103]. Also, in contrast to graphene that has a metallic behavior with zero bandgap, GO is a dielectric material with a large bandgap > 2 eV [88, 96] that yields both low linear light absorption and TPA in the telecom band, which are highly desirable for third-order nonlinear processes such as FWM and the Kerr effect that gives rise to SPM. [94] Further, there is the opportunity to enhance the overall nonlinear device performance even further through the combined use of GO films with p-i-n junctions to reduce the effects of TPA generated free carriers even further. Fig. 1(b) shows a schematic of an SOI nanowire waveguide integrated with a GO film. The fabrication of the SOI nanowire can be achieved via either deep ultraviolet photolithography or e-beam lithography followed by inductively coupled plasma etching, all of which are mature silicon device fabrication technologies [20, 104]. The GO film coating, with a thickness of ~2 nm per layer [82], can be achieved using solution-based methods that yield layer-by-layer film deposition [96, 97, 100]. As compared with the sophisticated transfer processes for other 2D materials such as graphene and TMDCs [105, 106], these coating methods enable transfer-free and conformal film coating, with high fabrication stability, repeatability, precise control of the film thickness (i.e., number of layers), and extremely good film attachment onto integrated photonic devices [94]. Precise control of the film length and coating position can be achieved by patterning the film with standard lithography and lift-off processes [99, 100]. This, together with the accurate control of the film thickness, allows the optimization of the Kerr nonlinear performance by adjusting the film thickness, length, and coating position.

Fig. 1(c-i) shows the schematic cross section of a hybrid waveguide with 1 layer of GO, while the corresponding transverse electric (TE) mode profile is shown in Fig. 1(c-ii). The interaction between the film and waveguide evanescent field excites the nonlinear optical response of the highly nonlinear GO film. Table I shows the definition of the parameters that we use to investigate the Kerr nonlinear optical performance of the GO-coated SOI nanowires, including the waveguide dimensions ($W$ and $H$), GO film parameters ($N$, $L_c$ and $L_0$), and pulsed laser parameters ($PE$ and $C_0$). Following our previous experimental measurements [82, 83, 100], the GO film thickness is assumed to be proportional to $N$, with a thickness of 2 nm per layer. We assume that the input pulse shape has a Gaussian profile:

$$A = \sqrt{P_0} \cdot exp[-\frac{1}{2}(1+iC_0)(\frac{t}{T_0})^2] \quad (1)$$

where $P_0$ is the pulse peak power, $C_0$ is the initial chirp, and $T_0$ is the half-width at $1/e$ intensity. The corresponding pulse energy ($PE$) can be described as

$$PE = P_0 \cdot T \quad (2)$$

where $T = 1.67 T_0$ is the pulse duration.

In the following sections, we first investigate the influence of waveguide geometry ($W$, $H$) and GO film thickness ($N$) on the linear and nonlinear loss of the hybrid waveguides in Section III, followed by their effect on the effective nonlinear parameter and nonlinear FOM in Section IV. In Section V, SPM-induced spectral broadening of optical pulses is investigated to illustrate the trade-off between optimizing the nonlinear FOM and minimizing linear loss. Using the results of Sections III and IV, we optimize the spectral broadening in GO-coated SOI nanowires by adjusting the device parameters such as waveguide geometry ($W$, $H$), layer number ($N$), pattern length ($L_c$), and coating position ($L_0$). Finally, we discuss the influence of loss, pulse chirp, material anisotropy, and the interplay between SA and SPM on the Kerr nonlinear performance in Section VI.

TABLE I
DEFINITIONS OF PARAMETERS OF WAVEGUIDES DIMENSION, GO FILM, AND PULSE LASER.

| Waveguide dimension | Height | Width |
|---|---|---|
| Parameters | $H$ | $W$ |
| GO film | GO layer number | Coating length | Uncoated length before GO segment |
| Parameters | $N$ | $L_c$ | $L_0$ |
| Pulse laser | Pulse energy (coupled into the waveguide) | Chirp |
| Parameters | $PE$ | $C_0$ |

. TABLE II
MATERIAL & WAVEGUIDE PARAMETERS USED IN OUR SIMULATIONS & ANALYSIS

| Material parameters [a] | Silicon | GO | Waveguide parameters [a] | SOI waveguide | Hybrid waveguide |
|---|---|---|---|---|---|
| Refractive index [b] | $n_{Si}$ : 3.48 [1] | $n_{GO}$ : 1.97 [47] | Linear propagation loss (m$^{-1}$) | $\alpha_{L\text{-}SOI}$ ($W$, $H$) | $\alpha_{L\text{-}hybrid}$ ($W$, $H$, $N$) |
| Extinction coefficient | $k_{Si}$ : 0 [32] | $k_{GO}(N)$ : 0.0079 – 0.0091 [32] | Effective mode area (m$^2$) | $A_{eff\text{-}SOI}$ ($W$, $H$) | $A_{eff\text{-}hybrid}$ ($W$, $H$, $N$) |
| Kerr coefficient (m$^2$/W) | $n_{2\text{-}Si}$ : 6 × 10$^{-18}$ [59] | $n_{2\text{-}GO}(N)$ : 1.22 × 10$^{-14}$ – 1.42 × 10$^{-14}$ [32] | Dispersion (s$^2$/m) | $\beta_{2\text{-}SOI}$ ($W$, $H$) | $\beta_{2\text{-}hybrid}$ ($W$, $H$, $N$) |
| TPA coefficient (m/W) | $\beta_{TPA\text{-}Si}$ : 5 × 10$^{-12}$ [59] | N/A [c] | Free carrier density (m$^{-3}$) | $N_{c\text{-}SOI}$ ($W$, $H$) | $N_{c\text{-}hybrid}$ ($W$, $H$, $N$) |
| FCA coefficient (m$^2$) | $\sigma$ : 1.45 × 10$^{-21}$ [59] | N/A [c] | TPA loss (m$^{-1}$) | $\alpha_{TPA\text{-}SOI}$ ($W$, $H$, $PE$) | $\alpha_{TPA\text{-}hybrid}$ ($W$, $H$, $PE$, $N$) |
| Carrier lifetime (s) | $\tau_c$ : 1 × 10$^{-9}$ [59] | N/A [c] | FCA loss (m$^{-1}$) | $\alpha_{FCA\text{-}SOI}$ ($W$, $H$, $PE$) | $\alpha_{FCA\text{-}hybrid}$ ($W$, $H$, $PE$, $N$) |
| SA coefficient (m$^{-1}$) | N/A [c] | $\alpha_{sat}(N)$ : 211 – 2099 [32] | SA loss (m$^{-1}$) | N/A [b] | $\alpha_{SA\text{-}hybrid}$ ($W$, $H$, $PE$, $N$) |
| Saturation intensity (W) | N/A [c] | $I_{sat}(N)$ : 0.06 – 0.24 [32] | Nonlinear parameter (W$^{-1}$m$^{-1}$) | $\gamma_{SOI}$ ($W$, $H$) | $\gamma_{hybrid}$ ($W$, $H$, $N$) |

[a] For simplification, the material and waveguide parameters in relevant equations are shown without the variables in the brackets.
[b] Here we show the refractive indices at 1550 nm, the same applies for other material parameters in this Table.
[c] The TPA of GO and the saturable absorption of silicon are neglected in our simulation

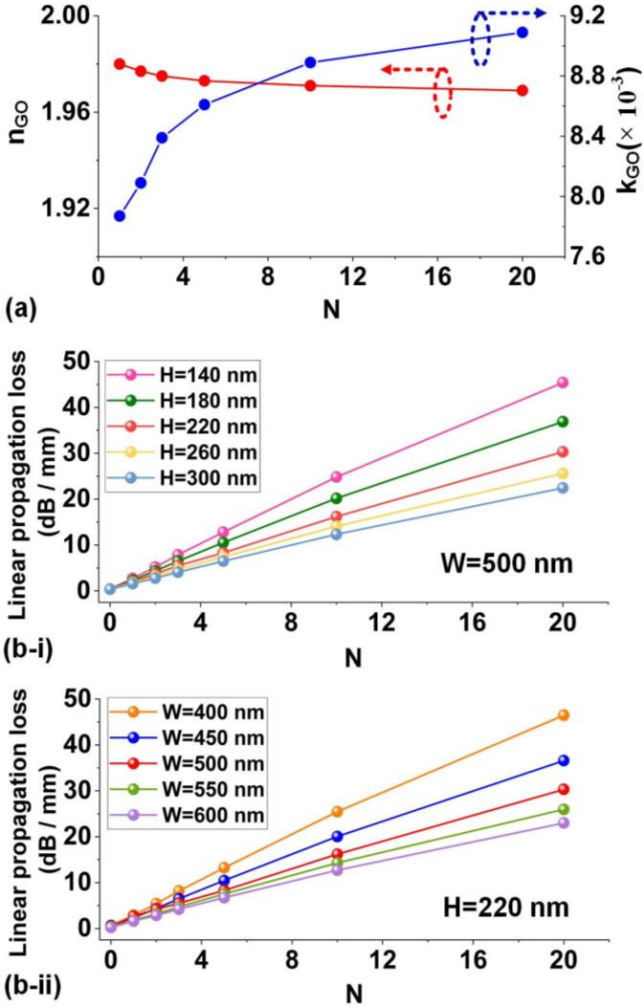

Fig. 2. (a) Refractive index $n_{GO}$ and extinction coefficient $k_{GO}$ of GO versus layer number $N$. (b) Linear propagation loss versus $N$ for GO-coated SOI nanowires with (i) various $H$ when $W$ = 500 nm and (ii) various $W$ when $H$ = 220 nm. The points at $N = 0$ correspond to the results for bare SOI nanowires

## III. LINEAR AND NONLINEAR LOSS

In this section, we investigate the linear and nonlinear loss of GO-coated SOI nanowires with different waveguide geometries ($W$, $H$) and GO film thickness ($N$). Table II summarizes the material parameters for silicon and GO as well as the waveguide parameters for the bare SOI nanowires and hybrid waveguides used our following simulations and analysis. For the waveguide parameters, we highlight their relationship with the physical parameters in Table I. The material parameters are independent of waveguide geometry, while the waveguide parameters are determined by not only the material parameters but also the waveguide geometry ($W$, $H$). Some of the waveguide parameters are also dependent on the GO film thickness ($N$) and pulse energy ($PE$). Fig. 2(a) shows the in-plane refractive index ($n_{GO}$) and extinction coefficient ($k_{GO}$) of GO (at 1550 nm) versus layer number $N$, measured by spectral ellipsometry. The $k_{GO}$ slightly increases with layer number $N$, varying from 0.0079 for $N = 1$ to 0.0091 for $N = 20$, mainly induced by scattering loss stemming from film unevenness and imperfect contact between the multiple GO layers. Note that the $k_{GO}$ is over two orders of magnitude lower than that of undoped graphene, highlighting its low linear light absorption and strong potential for high-performance nonlinear photonic devices. In principle, GO films with a bandgap > 2 eV should have negligible linear light absorption at telecom wavelengths, however, this is not the case for practical materials. We therefore infer that the linear loss of the GO films is mainly induced by light absorption from localized defects as well as scattering loss arising from film unevenness and imperfect contact between the different layers. Since these sources depend on the fabrication processes and are thus hard to quantitatively analyze, it is difficult to estimate the minimum linear loss that practical GO films can achieve. What is clear, however, is that there is still significant room to reduce the linear loss arising from these sources by optimizing the film fabrication processes. For the film refractive index $n_{GO}$, in contrast to the layer dependent $k_{GO}$, we assume that it is independent of the layer number $N$, consistent with the observation in Fig. 2(a). Figs. 2(b-i) and (b-ii) depict the linear propagation loss of the hybrid waveguides versus layer number ($N$), first for different waveguide heights ($H$) at a fixed width ($W$) and then for different widths ($W$) at a fixed height ($H$). The propagation loss of the hybrid waveguides was calculated using Lumerical FDTD commercial mode solving software with the $n_{GO}$ and $k_{GO}$ of layered films from Fig. 2(a). For the bare silicon waveguides ($N = 0$), we mainly considered the scattering loss arising from the sidewall roughness and used the classical theory in Ref. [107] to calculate the propagation loss of the bare SOI nanowires with different waveguide widths as follows:

$$\alpha_{L\text{-}SOI} = \frac{16\sigma^2}{\sqrt{2}\beta_0 W^4 n_{core}} gf \quad (3)$$

where $\alpha_{L\text{-}SOI}$ is the propagation loss coefficient, $\sigma$ is the standard deviation of the roughness, $\beta_0$ is the propagation constant in vacuum, $n_{core}$ is the core refractive index, $g$ and $f$ are parameters that can be calculated from $W$, waveguide propagation constant $\beta$, $n_{core}$, and cladding refractive index $n_{clad}$. For the variation of propagation loss with waveguide height $H$, we first used the effective index method in Ref. [108] to transfer the waveguide into an equivalent planar waveguide and then calculated the propagation loss based on Eq. (3). In our calculation, we assume $\sigma = 2$ nm, which is a typical value of average nanofabrication facilities such as ours and is consistent with the experimental propagation loss that we achieve. The calculated propagation loss varied from 2.2 dB/cm to 6.9 dB/cm for 25 different waveguide geometries (5 different $H \times$ 5 different $W$) considered in our simulations. We chose the transverse electric (TE) polarization because it supports an in-plane interaction between the evanescent field and film, which is much stronger than the out-of-plane interaction due to the large optical anisotropy of 2D materials [80, 100, 102]. In Figs. 2(b-i) and (b-ii), the propagation loss is seen to increase with layer number $N$ − a combined result of both increased $k_{GO}$ and GO mode overlap. The propagation loss decreases with waveguide height $H$ and width $W$, which shows the opposite trend to its change with layer number $N$, reflecting an increased GO mode overlap in SOI nanowires

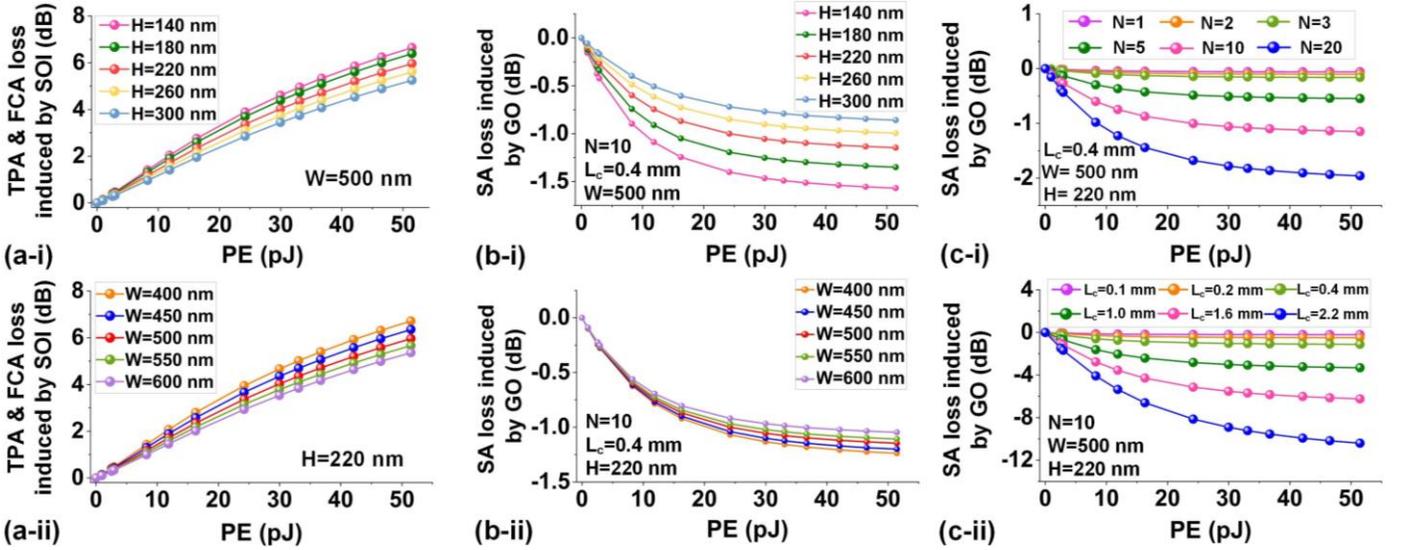

Fig. 3. (a) Power-dependent TPA and FCA loss induced by bare SOI nanowires. (b) Power-dependent SA loss induced by GO films when $N = 10$, $L_c = 0.4$ mm, and $L_0 = 1.3$ mm. In (a) and (b), (i) and (ii) show the results for SOI nanowires with various $H$ when $W = 500$ nm and various $W$ when $H = 220$ nm, respectively. (c) Power-dependent SA loss induced by GO films for (i) various $N$ when $L_c = 0.4$ mm and (ii) various $L_c$ when $N = 10$. In (c), $W = 500$ nm, $H = 220$ nm, and $L_0 = 1.3$ mm. In (a) – (c), the total length of the bare SOI nanowires is 3 mm.

that feature smaller waveguide dimensions.

Fig. 3(a) depicts the time-averaged nonlinear loss arising from TPA and free carrier absorption (FCA) of silicon versus pulse energy (*PE*) for the bare SOI nanowires, (i) for different heights (*H*) at a fixed width (*W*) and (ii) for different widths (*W*) at a fixed height (*H*). The pulsed laser parameters are: pulse duration $T = 3.9$ ps and initial chirp $C_0 = -0.3$ – taken from our previous experiments [82]. The *PE* varies from 0.38 pJ to 51.5 pJ, corresponding to a varied peak power from 0.1 W to 13.2 W. The TPA and FCA loss was calculated based on [109]:

$$\alpha_{TPA\text{-}SOI} + \alpha_{FCA\text{-}SOI} = \frac{\beta_{TPA\text{-}Si}}{A_{eff\text{-}SOI}} |A(z,t)|^2 + \sigma N_{c\text{-}SOI} \quad (4)$$

where $\beta_{TPA\text{-}Si} = 5 \times 10^{-12}$ m/W and $\sigma = 1.45 \times 10^{-21}$ m$^2$ are the TPA and FCA coefficients of silicon, respectively. These two parameters are determined by the intrinsic material properties, and so are independent of the waveguide geometry. In principle, they vary with the wavelength of light, however, in our case we neglect this since our pulse spectral width (< 10 nm) is much smaller than the dispersion bandwidths of these parameters. $A(z, t)$ is the slowly varying temporal envelope of the optical pulse along the waveguide (i.e., *z* axis). Note that Eq. (4) works for not only optical pulses but also for continuous-wave light. At a certain time within the pulse duration, the corresponding nonlinear loss was calculated based on Eq. (4), and then the nonlinear loss at different times was averaged to obtain the time-averaged nonlinear loss in Fig. 3(a). $A_{eff\text{-}SOI}$ is the effective mode area depending on the waveguide geometry and mode distribution. The variation of $A_{eff\text{-}SOI}$ (and so $N_{c\text{-}SOI}$) with waveguide geometry was considered in our following analysis. $N_{c\text{-}SOI}$ is the free carrier density given by [109]:

$$\frac{\partial N_{c\text{-}SOI}(z,t)}{\partial t} = \frac{\beta_{TPA\text{-}Si}}{2\hbar\omega} \cdot \frac{|A(z,t)|^4}{A_{eff\text{-}SOI}^2} - \frac{N_{c\text{-}SOI}(z,t)}{\tau_c} \quad (5)$$

where $\hbar$ is Planck's constant, $\omega$ is the angular frequency, and $\tau_c = \sim 1$ ns is the effective carrier lifetime. We neglect any free carrier effects in the GO films, and so the films do not significantly affect the carrier recombination time $\tau_c$ in the silicon waveguides. This is because GO has a large bandgap (typically > 2 eV) that is more than both the single- (0.8 eV) and two-photon (1.6 eV) energies at telecom wavelengths, which leads to a low efficiency of carrier generation due to either linear light absorption or TPA. When the pulse duration is much shorter than $\tau_c$, the $\tau_c$ term in Eq. (5) can be ignored as the generated free carriers do not have enough time to recombine within the pulse duration [106, 109]. In our analysis of nonlinear loss and SPM broadening, we considered a pulse train with a repetition rate of 60 MHz – the same that was used in our previous experiment [82]. Since the time duration between adjacent pulses (1 / 60MHz = ~16.7 ns) is much longer than the carrier recombination time ($\tau_c = 1$ ns), there is more than sufficient time for the carriers to recombine before the next pulse. The loss in Fig. 3(a) decreases with waveguide height *H* and width *W*, indicating a stronger TPA and FCA in SOI nanowires that have smaller waveguide dimensions.

As noted previously, the TPA and FCA of GO is very low at near-infrared wavelengths [96, 97] and so the nonlinear loss of GO is dominated by SA arising from the ground-state bleaching of the *sp*$^2$ domain with a typical energy gap of ~0.5 eV [92, 95, 110]. Fig. 3(b) shows the SA loss of the GO films for the hybrid waveguides with different waveguide geometries (*H* and *W*) but the same GO film parameters of $N = 10$ and $L_c = 0.4$ mm. The SA loss was calculated via [111, 112]:

$$\alpha_{SA\text{-}hybrid} = \alpha_{sat} / (1 + \frac{\eta|A|^2}{I_{sat}}) \quad (6)$$

where $\alpha_{sat}$ is the SA coefficient, $I_{sat}$ is the saturation intensity, and $\eta$ is the GO mode overlap. In our calculations, the layer dependent $\alpha_{sat}$ and $I_{sat}$ were obtained from experimental results [82] and $\eta$ was calculated via COMSOL Multiphysics. Here, $\alpha_{sat}$ and $I_{sat}$ are determined by the film properties. Given the broadband response of 2D layered GO films [100], the SA of GO was assumed to be wavelength independent. In Fig. 3(b), the SA loss becomes more significant as the waveguide height $H$ and width $W$ both decrease. This mainly results from an increase in $\eta$, slightly offset by a decrease in pulse energy $PE$ at the start of the GO coated segments (resulting from an increase in linear propagation loss in Fig. 2). Fig. 3(c) shows the SA loss versus pulse energy $PE$, for (i) different layer numbers $N$ at a fixed coating length of $L_c = 0.4$ mm and (ii) different coating lengths $L_c$ at a fixed layer number of $N = 10$. It is clear that SA in the films becomes more significant as the layer number $N$ and coating length $L_c$ both increase, reflecting more significant SA in the hybrid waveguides with thicker and longer films.

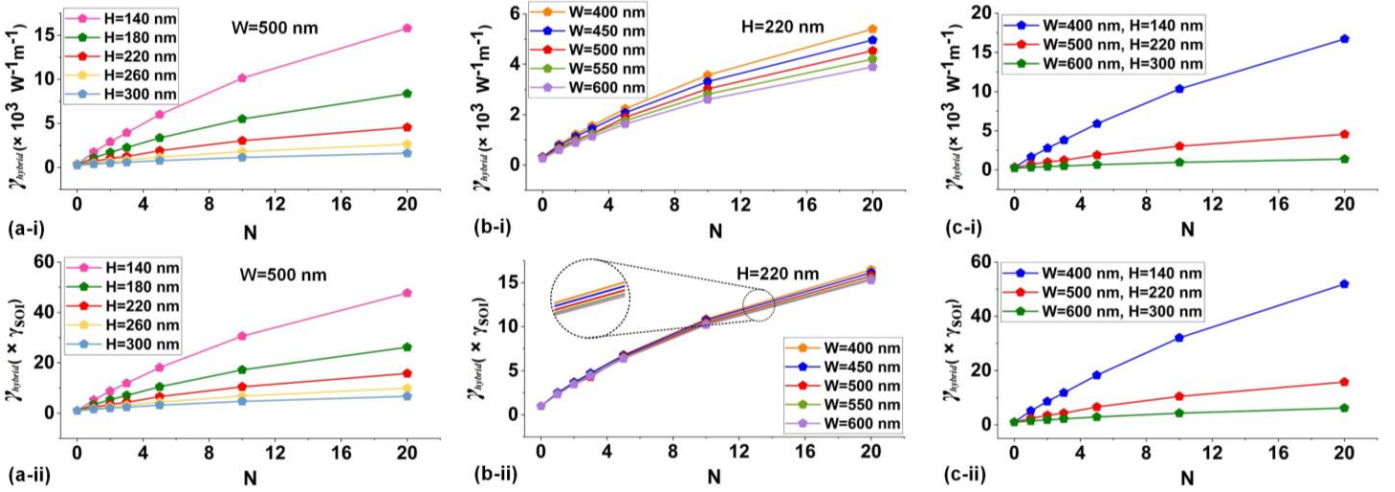

Fig. 4. Effective nonlinear parameter ($\gamma_{hybrid}$) versus $N$ for GO-coated SOI nanowires with (a) various $H$ when $W = 500$ nm, (b) various $W$ when $H = 220$ nm, and (c) the maximum, medium, and minimum waveguide dimensions. (i) shows the absolute $\gamma_{hybrid}$ values and (ii) shows the relative $\gamma_{hybrid}$ normalized to comparable bare SOI nanowires with the same waveguide geometries. The points at $N = 0$ correspond to the results for bare SOI nanowires.

## IV. NONLINEAR PARAMETER AND NONLINEAR FOM

In this section, we further investigate the influence of waveguide geometry ($W$ and $H$) and GO film thickness ($N$) on the effective nonlinear parameter and nonlinear FOM. Figs. 4(a) and (b) show the effective nonlinear parameter $\gamma_{hybrid}$ versus layer number $N$, first for 5 different waveguide heights $H$ at a fixed width $W$ and then for 5 different widths $W$ at a fixed height $H$. We plot the results for $\gamma_{hybrid}$ for both (i) the absolute value and (ii) the relative value normalized to bare SOI nanowires with the same geometries. Note that, since $\gamma_{SOI}$ for the bare SOI nanowires is geometry dependent, the absolute $\gamma_{hybrid}$ in (i) does not scale exactly as the relative $\gamma_{hybrid}$ in (ii). In Fig. 4(c), we also plot the corresponding results for hybrid waveguides with the maximum, intermediate, and minimum waveguide dimensions amongst the 25 considered waveguide geometries. The $\gamma$'s were calculated based on [97, 99]

$$\gamma_{hybrid} = \frac{2\pi}{\lambda_c} \frac{\iint_D n^2(x,y) n_2(x,y) S_z^2 dxdy}{\left[\iint_D n(x,y) S_z dxdy\right]^2} \quad (7)$$

where $\lambda_c$ is the pulse central wavelength, $D$ is the integral of the optical fields over different material regions, $S_z$ is the time-averaged Poynting vector calculated using COMSOL Multiphysics, $n(x, y)$ and $n_2(x, y)$ are the linear refractive index and $n_2$ profiles over the waveguide cross section, respectively. We used $n_2(x, y)$ rather than the more general third-order nonlinearity ($\chi^{(3)}$) because the pulse spectral width (< 10 nm) is much smaller compared to the dispersion bandwidth of $n_2$. The values of $n_2$ for silica and silicon used in our calculation are $2.60 \times 10^{-20}$ m$^2$/W [2] and $6 \times 10^{-18}$ m$^2$/W, respectively, with the latter obtained by fitting experimental results for the bare SOI nanowires [82], which agrees with other results [109]. The $n_{2\text{-GO}}$ for layered GO, obtained from experimental results [82], decreases from $1.44 \times 10^{-14}$ m$^2$/W for $N = 1$ to $1.24 \times 10^{-14}$ for $N = 20$. We suggest that this mainly results from an increase in inhomogeneous defects and imperfect contact in thicker films. We neglect any changes in $n_{2\text{-GO}}$ with pulse energy $PE$ and wavelength since these parameters vary by much less than what these changes are observed for.

In Figs. 4(a) and (b), $\gamma_{hybrid}$ increases with layer number $N$ and decreases with waveguide height $H$ and width $W$, showing a similar trend to the propagation loss in Figs. 2(a) and (b). This indicates that an increased mode overlap leads to both an increased Kerr nonlinearity as well as linear loss. In Fig. 4(c), when $N = 20$, $W = 400$ nm, and $H = 140$ nm, a high $\gamma_{hybrid}$ of 16711 W$^{-1}$m$^{-1}$ is obtained, which is ~52 times that of the associated bare SOI nanowire (with the same waveguide geometry) and ~4 times that of a comparable hybrid waveguide (with the same GO film thickness) with $W = 500$ nm and $H = 220$ nm. These results reflect the huge improvement in Kerr nonlinearity that can be obtained by not only introducing the GO films into SOI nanowires but by properly optimizing $\gamma_{hybrid}$ by engineering the GO mode overlap.

Based on the effective nonlinear parameter of the hybrid waveguides, we further investigate the effective nonlinear FOM ($FOM_{eff}$), which is widely used to quantitively characterize the trade-off between the Kerr nonlinearity and nonlinear loss [2]. Figs. 5(a) and (b) show the $FOM_{eff}$ (normalized to the FOM of silicon) versus layer number $N$, first for different waveguide heights $H$ at a fixed width $W$ and then for different widths $W$ with a fixed height $H$. The corresponding results for the hybrid waveguides with maximum, intermediate, and minimum waveguide dimensions are shown in Fig. 5(c). The $FOM_{eff}$'s were calculated by [1, 2]:

$$FOM_{eff}(N) = \frac{n_{2\text{-}eff}(N)}{\lambda_c \beta_{TPA\text{-}eff}(N)} \quad (8)$$

where $\beta_{TPA\text{-}eff}(N)$ is an effective TPA coefficient for the hybrid waveguide obtained by modeling the nonlinear loss arising from SA in the GO film (i.e., $\alpha_{SA\text{-}hybrid}$ - $\alpha_{sat}$) as being equivalent to an effective negative TPA and FCA loss. To obtain the $\beta_{TPA\text{-}eff}(N)$, we first calculated the overall nonlinear loss of the hybrid waveguides including both the TPA/FCA loss induced by silicon and the SA loss induced by GO. After that, we obtained the $\beta_{TPA\text{-}eff}$ by changing the TPA coefficient in Eqs. (4) and (5) to fit the overall nonlinear loss of the hybrid waveguides. Note that for a device with a patterned GO film, only the TPA coefficient for the hybrid segment was changed to obtain the corresponding $\beta_{TPA\text{-}eff}$. In Eq. (8), $n_{2\text{-}eff}(N)$ is the effective Kerr coefficient calculated from:

$$n_{2\text{-}eff}(N) = \frac{\lambda_c \gamma_{hybrid} A_{eff\text{-}hybrid}}{2\pi} \quad (9)$$

where $\gamma_{hybrid}$ is the effective nonlinear parameter in Fig. 4 and $A_{eff\text{-}hybrid}$ is the effective mode area of the hybrid waveguide. The $\beta_{TPA\text{-}eff}$'s are influenced by SA in the GO layers, which becomes more significant as the mode overlap increases. The SA decreases the overall absorption as the pulse energy $PE$ increases, which acts oppositely to TPA and results in the effective $\beta_{TPA\text{-}eff}$'s of the hybrid waveguides being smaller than that of comparable bare SOI nanowires having the same waveguide geometries.

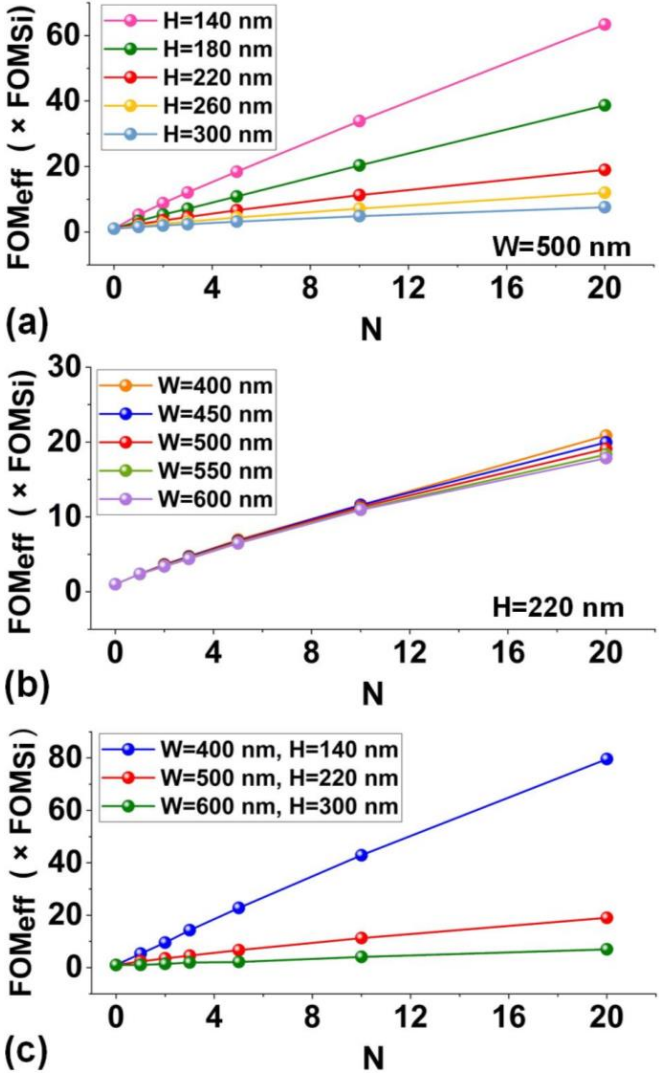

Fig. 5. $FOM_{eff}$ of the hybrid waveguides versus $N$ for GO-coated SOI nanowires with (a) various $H$ when $W = 500$ nm, (b) various $W$ when $H = 220$ nm, and (c) the maximum, intermediate, and minimum waveguide dimensions. The points at $N = 0$ correspond to the results for bare SOI nanowires.

As shown in Fig. 5(c), the increased $n_{2\text{-}eff}$ and reduced $\beta_{TPA\text{-}eff}$ yield a high $FOM_{eff}$ of 61 for $N = 20$, $W = 400$ nm, and $H = 140$ nm, which is ~79 times that of silicon and ~4 times that of a comparable hybrid waveguide (with the same GO film thickness) with $W = 500$ nm and $H = 220$ nm. Note that the FOM for bulk silicon reported in the literature varies by about a factor of 2. Our calculated FOM of ~0.7 agrees with that in Ref. [59], which is slightly higher than other reported values [2]. The $n_{2\text{-}Si}$ ($6 \times 10^{-18}$ m$^2$/W) and $\beta_{TPA\text{-}Si}$ ($5 \times 10^{-12}$ m/W) of silicon used in our calculations were obtained by fitting experimental results for the bare SOI nanowires [32]. In Figs. 5 (a) and (b), the effective FOM increases with layer number $N$ and decreases with waveguide height $H$ and width $W$, showing similar trends to the propagation loss in Fig. 2 and effective nonlinear parameter in Fig. 4. This indicates that the nonlinear FOM can be improved by increasing the GO mode overlap via reducing the waveguide geometry or increasing the GO film thickness.

## V. SPM-INDUCED SPECTRAL BROADENING OF OPTICAL PULSES

Although the nonlinear FOM (Eq. (8)) has been widely used to characterize the Kerr nonlinear optical performance of bulk materials [2, 113], it does not represent the full picture. The nonlinear optical performance of hybrid waveguides incorporating 2D materials (or indeed for any device), must factor in the effects of the linear propagation loss [94]. For GO-coated SOI nanowires, the increased mode overlap yields an increased nonlinear FOM, but at the expense of an increased linear loss that sometimes can be significant. In this section, we examine SPM-induced spectral broadening of optical pulses to illustrate this trade-off. We show that, in addition to the waveguide geometry and GO film thickness, other parameters such as the GO film length and coating position are also very important to optimize the Kerr nonlinear performance.

Fig. 6(a) compares the spectral broadening of optical pulses before and after propagating through both bare and GO-coated SOI nanowires, for various layer numbers $N$ at fixed $W = 500$ nm, $H = 220$ nm, and $L_c = 0.4$ mm. The total length of the bare

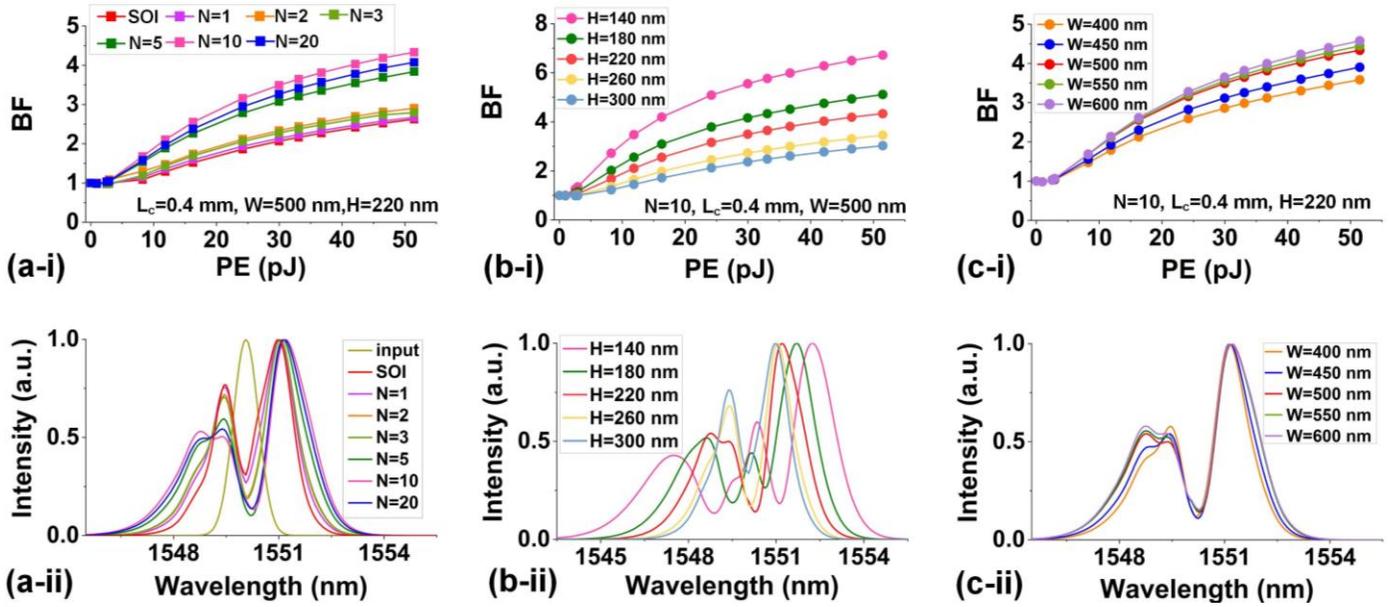

Fig. 6. (a) Spectral broadening of optical pulses before and after going through bare and GO-coated SOI nanowires with various $N$. (b) − (c) Spectral broadening of optical pulses after going through GO-coated SOI nanowires for various $H$ when $W = 500$ nm and various $W$ when $H = 220$ nm, respectively. In (a) − (c), (i) shows BFs versus $PE$ and (ii) shows the corresponding normalized spectra at $PE = 51.5$ pJ. In (a), $W = 500$ nm, $H = 220$ nm, and $L_c = 0.4$ mm. In (b) and (c), $N = 10$ and $L_c = 0.4$ mm. In (a) − (c), $L_0 = 1.3$ mm, and the total length of the bare SOI nanowires is 3 mm.

SOI nanowires is assumed to be 3 mm, as for previously fabricated devices [82]. To quantitatively analyze the spectral broadening, we use the broadening factor (BF) [82, 114, 115] defined as:

$$BF = \frac{\Delta\omega_{rms}}{\Delta\omega_0} \quad (10)$$

where $\Delta\omega_0$ and $\Delta\omega_{rms}$ are the root-mean-square (RMS) spectral width of the input and output optical spectra, respectively.

The spectral broadening was calculated using a split-step Fourier method to solve the nonlinear Schrödinger equation (NLSE) as follows [106, 109]:

$$\frac{\partial A}{\partial z} = -\frac{i\beta_{2\text{-hybrid}}}{2}\frac{\partial^2 A}{\partial t^2} + i\gamma_{hybrid}|A|^2 A - \frac{1}{2}i\sigma\mu N_{c\text{-hybrid}}A - \frac{1}{2}\alpha A \quad (11)$$

where $i = \sqrt{-1}$, $\beta_{2\text{-hybrid}}$ is the second-order dispersion coefficient, and $\beta_{2\text{-hybrid}}$ is the second-order dispersion coefficient. The variation of $\beta_{2\text{-hybrid}}$ with waveguide geometry and GO film thickness is considered in our simulation, although the results do not show a significant difference since in our case the physical length of the waveguides (i.e., 3 mm) is much shorter than their dispersion lengths (> 1 m). $\mu$ is the free carrier dispersion (FCD) coefficient of silicon, and $\alpha$ is the total loss including both the linear loss ($\alpha_{L\text{-hybrid}}$) in Fig. 2 and the nonlinear loss in Fig. 3, which can be expressed as:

$$\alpha = \alpha_{L\text{-hybrid}} + \alpha_{TPA\text{-SOI}} + \alpha_{FCA\text{-SOI}} + \alpha_{SA\text{-hybrid}} \quad (12)$$

In our calculations, the hybrid waveguides were separated into bare and GO-coated segments. Eq. (11) was solved for each segment, with the output from the previous segment used as the input to the following segment. Table III summarizes the flow to simulate and optimize the spectral broadening of optical pulses after going through the hybrid waveguide. To obtain the maximum BF, the five steps in Table III were repeated to calculate the BFs for the waveguides with different $W$, $H$, $N$, and $L_c$. In Fig. 6(a), the hybrid waveguides show more significant spectral broadening than the bare SOI nanowires, with the maximum spectral broadening being achieved for an intermediate number of layers, $N$ = 10. This results from the enhanced Kerr nonlinearity of the hybrid waveguides, balanced with the increased linear loss.

TABLE III
FLOW TO SIMULATE AND OPTIMIZE SPECTRAL BROADENING OF OPTICAL PULSES AFTER GOING THROUGH THE HYBRID WAVEGUIDES

| Step | Aim | Method & Theory | Used parameters | Figures |
|---|---|---|---|---|
| 1 | Calculation of linear propagation loss $\alpha_{L\text{-SOI}}, \alpha_{L\text{-hybrid}}$ | Lumerical FDTD | MP [a]: $n_{Si}, n_{GO}, k_{Si}, k_{GO}$<br>PP [b]: $W, H, N$ | Fig. 2 |
| 2 | Calculation of nonlinear loss $\alpha_{TPA}, \alpha_{FCA}, \alpha_{SA}$ | MATLAB<br>Eqs. (4) – (6) | MP: $\beta_{TPA\text{-Si}}, \sigma, \tau_c, \alpha_{sat}, I_{sat}$<br>PP: $W, H, N, PE, L_c$<br>WP [c]: $N_c, A_{eff}$ | Fig. 3 |
| 3 | Calculation of nonlinear parameter $\gamma_{SOI}, \gamma_{hybrid}$ | COMSOL<br>Eq. (7) | MP: $n_{2\text{-SOI}}, n_{2\text{-GO}}$<br>PP: $W, H, N$<br>WP: $A_{eff\text{-SOI}}, A_{eff\text{-hybrid}}$ | Fig. 4 |
| 4 | Simulation of broadened spectra after going through the waveguides | MATLAB<br>Eqs. (11) – (12) | Results in Step 1 – 3 | Figs. 6(a-ii) – 6(c-ii) |
| 5 | Calculation of broadening factor (BF) [d] | MATLAB<br>Eq. (10) | Result in Step 4 | Figs. 6(a-i) – 6(c-i) Fig. 8 |

[a] MP: material parameters in Table II.
[b] PP: physical parameters in Table I.
[c] WP: Waveguide parameters in Table II.
[d] To obtain the maximum BF, Step 1 – 5 were repeated to calculate the BFs for the waveguides with different $W, H, N$, and $L_c$.

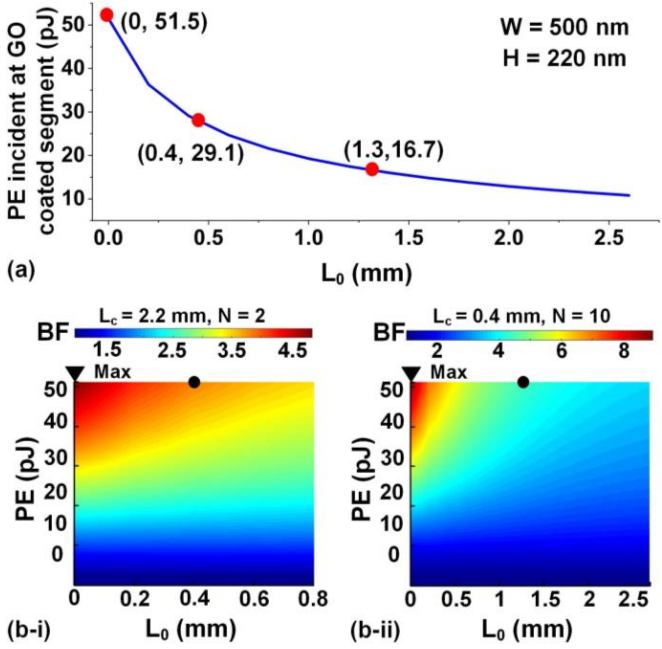

Fig. 7. (a) $PE$ incident at the GO-coated segment versus $L_0$. (b) BF of GO-coated SOI nanowires versus $L_0$ and $PE$ when (i) $L_c$ = 2.2 mm, $N$ = 2 and (ii) $L_c$ = 0.4 mm, $N$ =10. The black points mark the BFs of 3.75 at $L_0$ = 0.4 mm, $PE$= 51.5 pJ in (i) and 4.34 at $L_0$ = 1.3 mm, $PE$= 51.5 pJ in (ii). In (a) and (b), $W$ = 500 nm and $H$ = 220 nm.

Figs. 6(b) and (c) show the spectral broadening of optical pulses after propagation through bare and GO-coated SOI nanowires, for different waveguide geometries ($H$ and $W$) but with the same GO film parameters ($N$ = 10, $L_c$ = 0.4 mm, and $L_0$ = 1.3 mm). The spectral broadening becomes more significant as $H$ decreases, due to the significantly increased Kerr nonlinearity arising from the increased GO mode overlap, which dominates for the relatively short coating length considered here (i.e., $L_c$ = 0.4 mm). On the other hand, the spectral broadening increases with $W$, showing the opposite trend to that of the $FOM_{eff}$ in Fig. 5. This could reflect the fact that the linear loss can become a limiting factor for the nonlinear performance of the hybrid waveguides – otherwise the maximum spectral broadening would have been achieved for the smallest $W$ where the $FOM_{eff}$ is the greatest. We also note that the broadened spectra exhibits a slight asymmetry, which is mainly induced by the FCA and FCD in silicon [109].

Fig. 7(a) shows the pulse energy $PE$ incident at the GO-coated segment as a function of coating position $L_0$. The $PE$ decreases super-linearly with $L_0$, mainly induced by the super-linear increase in TPA and FCA of silicon. Fig. 7(b) shows the BFs versus coating position $L_0$ and pulse energy $PE$ when (i) $L_c$ = 2.2 mm, $N$ = 2 and (ii) $L_c$ = 0.4 mm, $N$ = 10. As expected, the spectral broadening becomes more significant as the $PE$ increases, due to the increased nonlinear efficiency. On the other hand, the BFs decrease with $L_0$, with the maximum values (BF = 4.8 in Fig. 7(b-i) and BF = 9.1 in Fig. 7(b-ii)) being achieved at $L_0$ = 0 mm, where there is a maximum $PE$ at the start of GO-coated segments. As marked in Figs. 7(b-i) and (b-ii), the BFs are 3.75 when $L_0$ = 0.4 mm, $PE$ = 51.5 pJ and 4.34 when $L_0$ = 1.3 mm, $PE$ = 51.5 pJ, showing good agreement with our previous experiments results [82]. In our previous experiments [82], windows were opened in the middle of the 3-mm-long SOI nanowires, resulting in $L_0$ = 0.4 mm and $L_0$ = 1.3 mm for the devices with 2.2-mm-long and 0.4-mm-long opened windows. This was mainly out of consideration of device coupling loss – silica-clad waveguide regions were introduced between the inverse taper couplers and the opened windows to increase the coupling efficiency. Note that the difference induced by $L_0$ gets smaller for lower propagation loss of the bare waveguides, being much lower for Hydex and SiN waveguides [83, 97] versus SOI nanowires studied here.

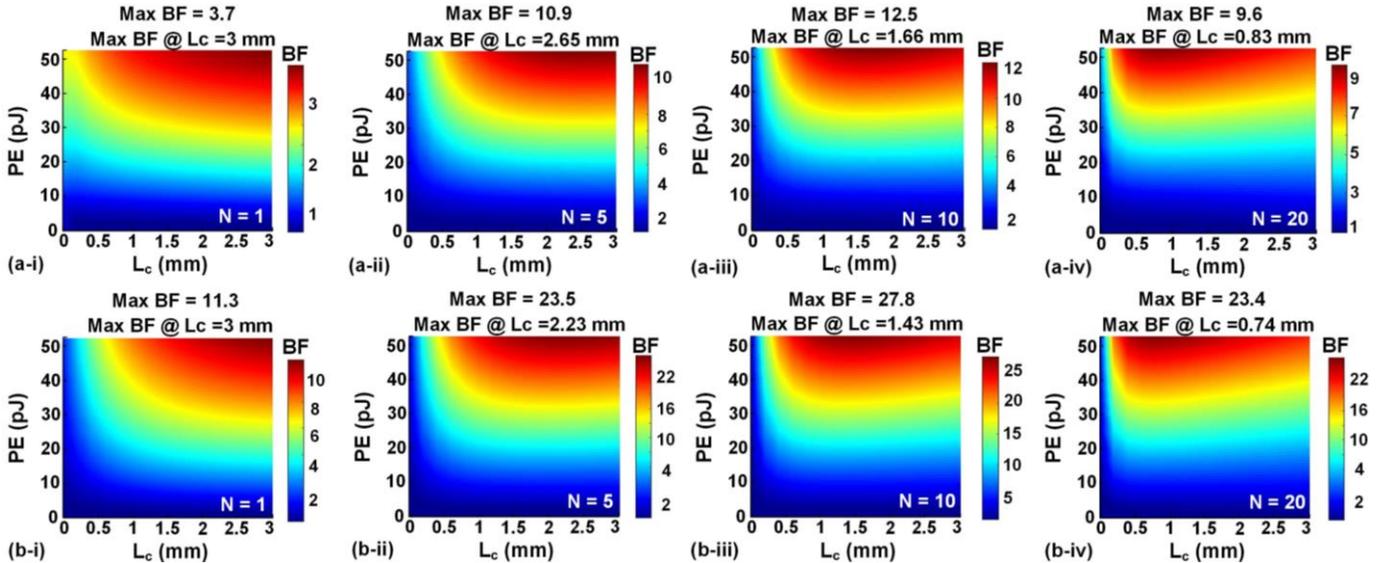

Fig. 8. SPM BF of GO-coated SOI nanowires versus $L_c$ and $PE$ when (a) $W$ = 500 nm, $H$ = 220 nm and (b) $W$ = 600 nm, $H$ = 140 nm. (i) – (iv) show the results for $N$ = 1, 5, 10, and 20, respectively. In (a) and (b), $L_0$ = 0 mm and when $L_c$ = 0 mm the BFs correspond to the results for the bare SOI nanowires.

Fig. 8(a) shows the BFs versus coating length $L_c$ and pulse energy $PE$ when (i) $N = 1$, (ii) $N = 5$, (iii) $N = 10$, and (iv) $N = 20$, respectively, with the other device parameters kept constant ($L_0 = 0$ mm, $W = 500$ nm, and $H = 220$ nm). The corresponding results for the bare SOI nanowires (when $L_c = 0$ mm) are also shown for comparison, and the maximum BFs (at 51.5 pJ) for the 500 nm × 220 nm and 600 nm × 140 nm bare SOI nanowires are ~2.6 and ~2.8, respectively. The maximum BFs are achieved for $L_c = 3$ mm, 2.65 mm, 1.66 mm, and 0.83 mm, respectively, which shifts towards shorter lengths as $N$ increases, following the trend also seen with the layer number $N$ in Fig. 6(a). This reflects the fact that the enhancement in the Kerr nonlinearity dominates for hybrid waveguides with relatively small $L_c$ and $N$, while the influence of the loss increase become more significant as $L_c$ and $N$ increase. A maximum BF of 12.5 is achieved when $N = 10$ and $L_c = 1.66$ mm, reflecting that there is still room for improvement on the basis of the maximum BF in Fig. 7(b) (i.e., 9.1) by optimizing the GO film length. Fig. 8(b) depicts the corresponding results for the hybrid waveguides with $W = 600$ nm and $H = 140$ nm, which shows the best spectral broadening among 25 different waveguide geometries considered in our study. A maximum BF of 27.8 is achieved when $N = 10$ and $L_c = 1.43$ mm, which is ~2.2 times higher than the maximum BF in Fig. 8(a) and more than 6 times higher than previous experiments [82], reflecting the potential for improvement by optimizing the waveguide geometry.

## VI. DISCUSSION

In this section, we first investigate the influence of loss, both linear and nonlinear, on the Kerr nonlinear performance. In order to compare with the previous experimental results and highlight any differences, we perform the simulations based on the SOI nanowire with a cross section of $W = 500$ nm and $H = 220$ nm. Fig. 9(a) shows the loss for optical pulses after propagation through hybrid waveguides, factoring in $\alpha_{L\text{-}hybrid}$, $\alpha_{L\text{-}hybrid}$ and $\alpha_{TPA\text{-}SOI}$, the overall loss excluding $\alpha_{SA\text{-}hybrid}$, and the total overall loss, calculated from Eqs. (4) – (6). The corresponding BFs and pulse spectra calculated from Eqs. (10) – (12) are shown in Figs. 9(b) and (c), respectively. The loss remains constant when considering only $\alpha_{L\text{-}hybrid}$ but increases with pulse energy $PE$ when including $\alpha_{TPA\text{-}SOI}$ and $\alpha_{FCA\text{-}SOI}$. We neglect any variation in $\alpha_{L\text{-}hybrid}$ with $PE$ for the hybrid waveguides, due to the much weaker photo-thermal changes of GO induced by picosecond optical pulses, with much lower average power than the continuous-wave pump for FWM [83, 99]. After including $\alpha_{SA\text{-}hybrid}$, the overall loss decreases, enhancing the SPM and spectral broadening. In Fig. 9(c), although the intrinsic TPA itself leaves the pulse spectrum symmetric, the resulting FCA makes it considerably asymmetric.

While the linear loss of the layered GO films does pose a limitation for the Kerr nonlinear performance of the hybrid waveguide, as mentioned, it can be further reduced, and any reduction resulting from optimizing the film fabrication processes would improve the nonlinear performance. Fig. 9(d) shows the linear propagation loss and BF of the hybrid waveguides versus GO extinction coefficient $k_{GO}$. For the GO film with the state-of-the art $k_{GO} = 0.0089$ (for $N = 10$), the corresponding propagation loss and BF are 15.8 dB/mm and 4.34, respectively, in good agreement with experiments [82]. When $k_{GO}$ decreases to 0.0008, the BF increases to 6.30 – a factor of ~1.5 higher than the BF for $k_{GO} = 0.0089$. Note that the GO coating length here ($L_c = 0.4$ mm) is relatively short, and so the influence of the loss of the GO films is not very significant. For the hybrid waveguide in Fig. 8 (b-iii) with a longer GO coating length of $L_c = 1.43$ mm, the BF increases from 27.8 (for $k_{GO} = 0.0089$) to 70 (for $k_{GO} = 0.0008$), highlighting the strong potential for improving the nonlinear performance by reducing the GO film linear loss.

Fig. 10(a) plots the mode overlap with the GO film ($\eta$) versus layer number $N$. Most of the power is confined to the SOI nanowires rather than the GO film (< 6% for $N = 20$), mainly due to the larger cross sectional area of the SOI nanowires compared to the ultrathin 2D GO films. In Fig. 10(a), we show the ratio of the power in the GO on both sidewalls to the power in entire film, which is < 3% and decreases with $N$. This indicates that the TE mode overlap with GO on both waveguide sidewalls is negligible compared with that for GO on the waveguide top, and so we use the in-plane $n_{2\text{-}GO}$ of GO (corresponding to TE polarization) in our calculations, neglecting any anisotropy in $n_{2\text{-}GO}$ of the 2D layered GO films.

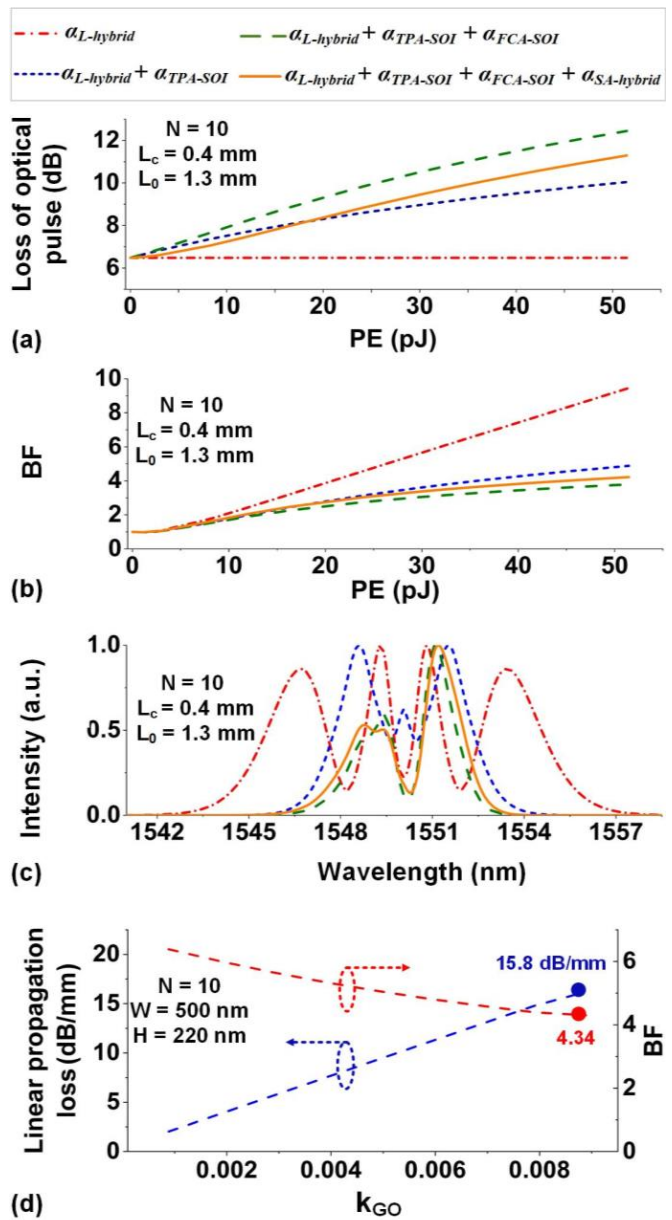

Fig. 9. (a) Loss of optical pulses after going through GO-coated SOI nanowires when considering $\alpha_{L\text{-}hybrid}$ (dotted curves), $\alpha_{L\text{-}hybrid}$ and $\alpha_{TPA\text{-}SOI}$ (short-dashed curves), overall loss except for $\alpha_{SA\text{-}hybrid}$ (dashed curves), and overall loss (solid curves). (b) BFs calculated based on the loss in (a). (c) Normalized spectra at $PE = 51.5$ pJ. (d) The linear waveguide propagation loss and BF of GO-coated SOI nanowires versus extinction coefficient $k_{GO}$. The points at $k_{GO} = 0.0089$ refer to the state-of-the-art values in Ref. [32]. In (a) – (d), the device parameters are $W = 500$ nm, $H = 220$ nm, $N = 10$, $L_c = 0.4$ mm, and $L_0 = 1.3$ mm.

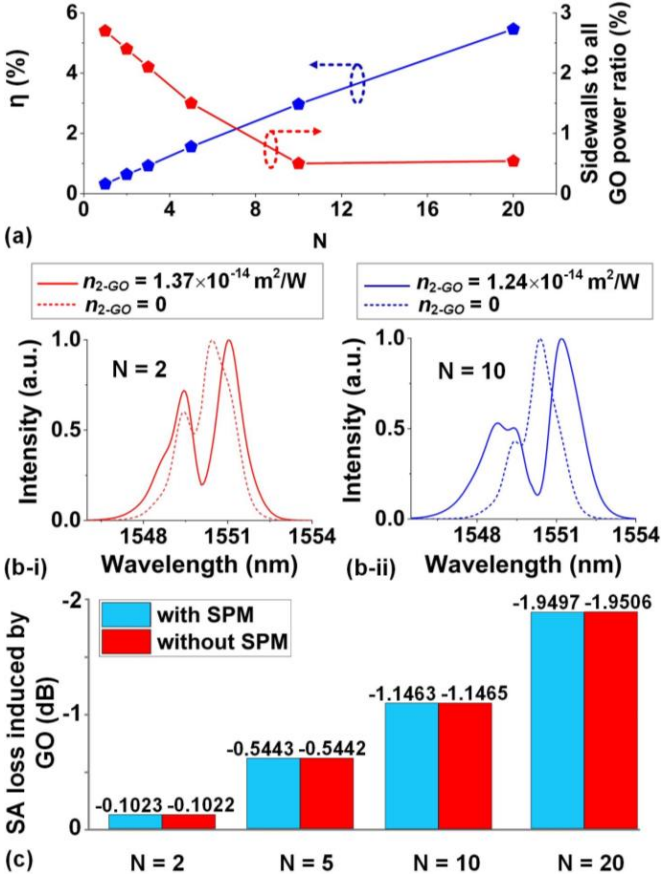

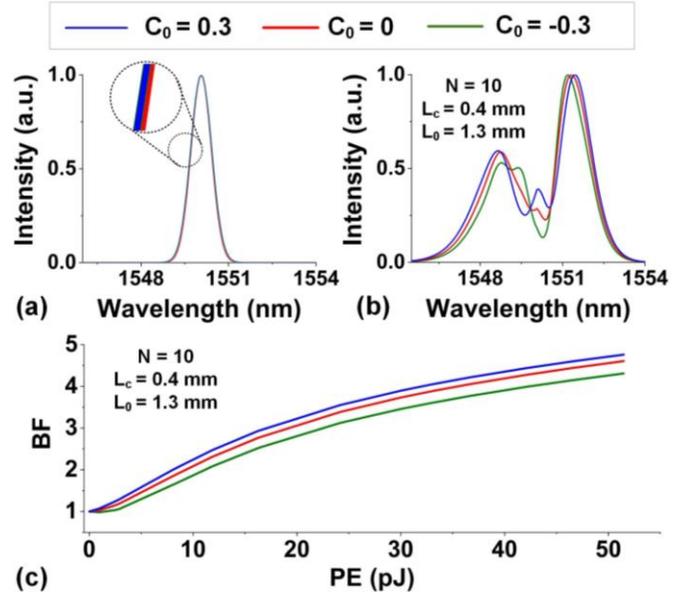

Fig. 11. Spectral broadening of optical pulses before and after going through GO-coated SOI nanowires for various $C_0$. (a) Input pulse spectra. (b) Normalized spectra at $PE$ = 51.5 pJ. (c) BFs versus $PE$. In (b) and (c), $W$ = 500 nm, $H$ = 220 nm, $N$ = 10, $L_c$ = 0.4 mm, and $L_0$ = 1.3 m.

Fig. 10. (a) GO mode overlap and ratio of power in GO coated on both sidewalls to that in all GO material regions versus $N$. (b) Comparison of optical pulse spectra after going through GO-coated SOI nanowires with (i) $N$ = 2 when $n_{2\text{-}GO}$ = 0 and $n_{2\text{-}GO}$ = 1.37 × 10$^{-14}$ m$^2$/W (ii) $N$ = 10 when $n_{2\text{-}GO}$ = 0 and $n_{2\text{-}GO}$ = 1.24 × 10$^{-14}$ m$^2$/W. (c) Comparison of SA loss of the hybrid waveguides with and without considering spectral broadening induced by GO. In (a), (b), and (c), $W$ = 500 nm, $H$ = 220 nm, $L_c$ = 0.4 mm, and $L_0$ = 1.3 mm.

Fig. 10(b) shows the optical pulse spectra after propagation through the hybrid waveguides with $N$ = 2 and $N$ = 10, respectively. The solid curves show the results based on the experimentally measured $n_{2\text{-}GO}$ (which varies slightly with $N$) [82], whereas the dashed curves were calculated neglecting the contribution of the GO film to the SPM (i.e., assuming $n_{2\text{-}GO}$ = 0). Fig. 10(c) compares the corresponding SA loss of the hybrid waveguides. The maximum difference between them is < 0.2%, reflecting that the influence of SPM on SA is negligible. This is mainly because the total length of the hybrid waveguides (3 mm) was much shorter than the dispersion length (> 1 m), and so any change in the pulse spectrum induced by SPM did not significantly affect the temporal pulse shape [114].

Fig. 11 compares the SPM performance for input optical pulses versus chirp parameter $C_0$. The pulse width $T$ as a function of $C_0$ is given by:

$$T = \frac{T_0}{(1+C_0^2)^{1/2}} \quad (13)$$

In Fig. 11(a), the input pulse spectra for the same absolute value of chirp (0.3) but with opposite signs overlap each other, in agreement with Eq. (13). The corresponding pulse spectra and BFs are shown in Figs. 11(b) and (c), respectively. Compared with unchirped pulses (i.e., $C_0$ = 0), the spectral broadening increases when $C_0 > 0$ but decreases when $C_0 < 0$, since the positive chirp induced by SPM adds to a positive $C_0$, while it is offset by a negative $C_0$. Note that our simulations in Figs. 6–10 were performed with $C_0$ = -0.3 in accordance with the pulsed laser used in our previous experiments [82].

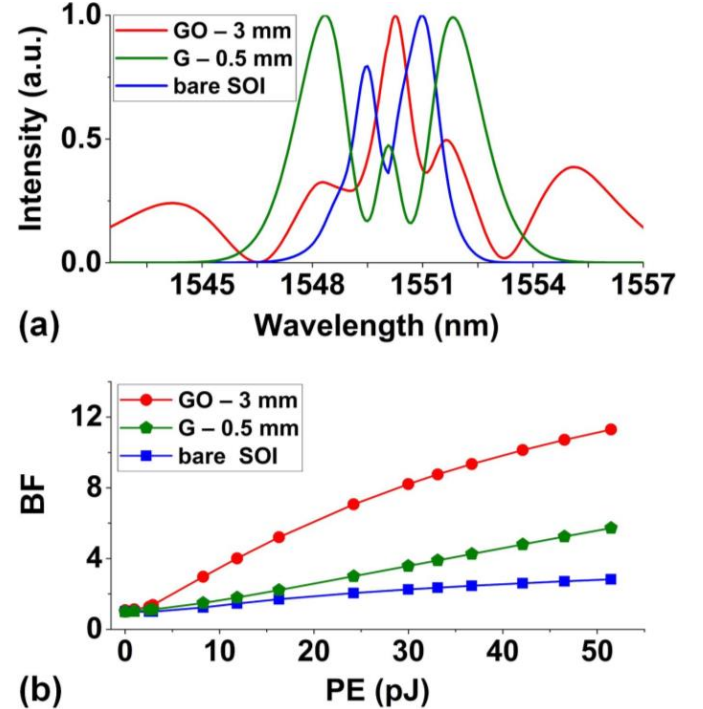

Fig. 12. Spectral broadening of optical pulses after going through a bare SOI nanowire and hybrid SOI nanowires conformally coated with monolayer GO and graphene. (a) Normalized output spectra at $PE$ = 51.5 pJ. (b) BFs versus $PE$. In (a) and (b), $W$ = 600 nm, $H$ = 140 nm, $N$ = 1, $L_0$ = 0 mm, $C_0$ = -0.3, and the total length of the SOI nanowire is 3 mm. The film coating lengths for the GO and graphene hybrid waveguides are 3 mm and 0.5 mm, respectively. G: graphene.

Finally, we compare the spectral broadening of optical pulses after going through a bare SOI nanowire and hybrid SOI nanowires conformally coated with monolayer GO and graphene in Fig. 12. Except for the film coating length and thickness, the other parameters are kept the same as $W$ = 600 nm, $H$ = 140 nm, $N$ = 1, and $L_0$ = 0 mm. We chose doped graphene with a lower linear loss than undoped graphene in our simulations, and the material parameters we used are $n_{graphene}$ = 2.8 [97], $k_{graphene}$ = 0.5 [116], $n_{2\text{-}graphene}$ = -1.1 ×10$^{-13}$ m$^2$/W [117], and monolayer film thickness = 1 nm [80]. For the GO hybrid waveguide, we choose the optimized coating length of $L_c$ = 3 mm in Fig. 8(b-i), where a maximum BF of ~11.3 is achieved at $PE$ = 51.5 pJ. For the graphene hybrid waveguide, we show the result for the optimized coating length of 0.5 mm, where a maximum BF of ~5.7 is achieved at $PE$ = 51.5 pJ. As can be seen, both GO and graphene hybrid waveguides show enhanced spectral broadening as compared with the bare SOI nanowire (with a maximum BF of ~2.8 at $PE$ = 51.5 pJ), and the GO hybrid waveguide exhibits even better spectral broadening than the graphene hybrid waveguide, reflecting its high Kerr nonlinear optical performance. Note that we did not account for the TPA of graphene in our calculation. In principle, pure graphene with a zero bandgap should have very strong TPA, while both TPA and SA behaviors have been reported for practical graphene [112, 118]. For practical graphene with TPA, the BF for the graphene hybrid waveguide would be lower than the simulation result in Fig. 12. Finally, we have recently also performed a theoretical analysis and design optimization for GO films in silicon nitride waveguides [119].

## VII. CONCLUSON

In summary, we theoretically investigate and optimize the Kerr nonlinear optical performance of SOI nanowires integrated with 2D layered GO films. Detailed analysis of the influence of waveguide geometry and GO film thickness on the propagation loss, nonlinear parameter, and nonlinear FOM is performed. We show that the effective nonlinear parameter and nonlinear FOM can be increased by up to ~52 and ~79 times relative to bare SOI nanowires, respectively. To examine the trade-off between increasing the Kerr nonlinearity and minimizing linear loss, we consider SPM-induced spectral broadening of optical pulses. We show that a high BF of 27.8 can be achieved by properly balancing this trade-off, more than a factor of 6 higher than what has been achieved experimentally. Finally, the role of pulse chirp, material anisotropy, and the interplay between SA and SPM in SPM-induced spectral broadening is also investigated, together with performance comparison of GO-coated and graphene-coated SOI nanowires. These results highlight the significant potential of GO films to enhance the Kerr nonlinear optical performance of SOI nanowires for practical applications.